\newcommand{\To}{\longrightarrow}

\newcommand{\RR}{\left( r \vert r' \right)}
\newcommand{\VT}{\tilde{V}}
\newcommand{\TV}{\VT}
\newcommand{\TTV}{\tilde{\TV}}
\newcommand{\bdelta}{\bar{\delta}}
\newcommand{\XY}{ \left ( x y \vert x y \right ) }
\newcommand{\xy}{ \left ( x y \vert x' y' \right ) }
\newcommand{\IM}{\hat{I}_{\vert m \vert}}

\documentclass[preprintnumbers,a4paper,amsfonts,amsmath,runinaddress]{revtex4}
\usepackage{graphicx}
\usepackage{longtable}
\usepackage{rotating}
\usepackage[usenames]{color}
\begin{document}



\pagestyle{empty}

\title{Semi-classical Expansion of the Smooth Part of the Density of States, with Application to the Hydrogen Atom in Uniform Magnetic Field}
\author{Herv\'e Kunz} \email{herve.kunz@epfl.ch}
\author{Thierry Sch\"upbach} \email{thierry.schuepbach@a3.epfl.ch}
\affiliation{Institute of Theoretical Physics, Swiss Federal Institute of Technology EPFL, Station 3, CH-1015 Lausanne, Switzerland}
\author{Marc-Andr\'e Dupertuis} \email{marc-andre.dupertuis@epfl.ch}
\affiliation{Laboratory of Quantum Optoelectronics, Swiss Federal Institute of Technology EPFL,
Station 3, CH-1015 Lausanne, Switzerland }

\keywords{Quantum Chaos, semi-classical limit, hydrogen atom}
\pacs{05.45.+b, 05.40.+j}
\date{April $30^{th}$, 2007 }


\begin{abstract}
We give a method to compute the smooth part of the density of states in a semi-classical expansion, when the Hamiltonian contains a Coulomb potential and non-cartesian coordinates are appropriate. We apply this method to the case of the hydrogen atom in a magnetic field with fixed $z$-component of the angular momentum. This is then compared with numerical results obtained by a high precision finite element approach. The agreement is excellent especially in the \emph{chaotic} region of the spectrum. The need to go beyond the Thomas-Fermi model is clearly established. 
\end{abstract}
\maketitle

Studies in quantum chaos require a knowledge of the density of states of a given system in the semi-classical limit. This quantity is decomposed into two different parts, a smooth one and an oscillating one, related to periodic orbits of the corresponding classical system.
A good knowledge of the smooth part is needed to unfold the spectrum before determining its statistical behavior. Simple examples show how a bad unfolding deteriorates the statistics \cite{Gomez2002}. \\
One of the standard tools to determine the smooth part of the density of states consists in computing its Laplace transform, i.e. the partition function in the semi-classical limit where a dimensionless $\hbar$ tends to zero but the temperature remains fixed. Quite often the partition function is expressed by a functional integral of the Feynman-Kac type \cite{Grosche1998}. But such an approach fails if the Hamiltonian contains a Coulomb potential. One way out of this difficulty is not to use cartesian coordinates, but more appropriate ones. Unfortunately it is difficult to work with a functional integral in non-cartesian coordinates despite progress made in some specific cases \cite{Grosche1998}. \\
We have therefore considered another approach based on the decomposition of the Hamiltonian into a differential operator and a multiplication operator. The differential operator is proportional to a dimensionless $\hbar^2$ and its propagator is supposed to be known exactly in some non-cartesian coordinates. This formulation of the problem is particularly well adapted to the case where the multiplication operator contains a Coulomb potential, since in this case the use of semi-parabolic coordinates suppresses the Coulomb singularity at the origin. There is however a price to be paid in the sense that the energy appears in the Hamiltonian. Nevertheless we have been able to compute the smooth part of the density of states up to order $\hbar^0$. We have applied these formulas to the case of the hydrogen atom in a uniform magnetic field, which is a paradigmatic model in the study of quantum chaos \cite{Friedrich1989, Delande1984}. In this case we work in the fixed $L_z$ vector. To the best of our knowledge these results are new.
\\
We then compared the results with numerical energy levels obtained by means of a high precision finite element approach to the problem of the hydrogen atom in a uniform magnetic field. Details of the numerical method will be given elsewhere.
The comparison of the analytical results with the numerical ones shows a very good agreement except in the lower part of the spectrum where the quasi-degeneracies specific to the hydrogen atom would require a different treatment. The comparison demonstrates the need to go beyond the Thomas-Fermi type part of the density of states in order to satisfactorily account for the numerical results. \\
Our approach suggests that it should also be possible to derive the oscillating part of the density of states in a satisfactory way. It is probably related to the periodic orbits in the way given in the literature \cite{Friedrich1989} but to the best of our knowledge, the usual derivation are not appropriate to the case of the hydrogen atom in a magnetic field.

\section{Semi-classical expansion of the smooth part of the density of states: A General Method}
A problem of great interest, and which has often been treated consists in finding an asymptotic expansion for the
density of states of an hamiltonian $H$, in terms of dimensionless parameter $\delta$ corresponding to Planck constant.

One way to achieve this is to find an asymptotic expansion for its Laplace transform, which is the partition function $Z(t)$ given by
\begin{equation} \label{partition}
Z(t) = \text{Tr } e^{-t H}
\end{equation}
and to go back to the density of states by the inverse Laplace transform of $Z(t)$. The following difficulty appears however in this program:
$t$ can be very large compared to $\delta^{-1}$ for example. If we ignore this difficulty however by considering that $t$ is fixed and $\delta$ is small, we will get an expansion for the so-called smooth part of the density of states and will miss the oscillating part discovered by Balian-Bloch \cite{Balian1970, Balian1972} and Gutzwiller \cite{Gutzwiller1990}. They are associated to terms like $\exp{- t \delta^{-1}}$ in the expansion of $Z(t)$. This is the strategy we will follow.

We will here consider the case where the hamiltonian is of the form 
\begin{equation} \label{Hamiltonian}
H = \delta^2 H_0 + V
\end{equation} 
and the propagator
\begin{equation} \label{propagator}
U_t \RR =  \left [ e^{-t \left ( \delta^2 H_0 + V \right ) } \right ] \RR
\end{equation}
$H_0$ being a differential operator and $V$ a multiplication operator. Our idea is the following. Consider that in the propagator given by Eq.~\ref{propagator}, $r$ is fixed and consequently so is $V(r)$, we can then rewrite the propagator in the form
\begin{equation}
U_t \RR = e^{-t V(r)} \, W_t \RR
\end{equation}
where 
\begin{equation}
W_t \RR = \left [ e^{-t \left ( \delta^2 H_0 + B \right )} \right ] \RR
\end{equation}
$B$ being a multiplication operator defined by 
\begin{equation}
\left ( B \psi \right ) (r') = \left [V(r') - V(r) \right ] \psi(r').
\end{equation}

 We can now use the iterative representation of $W_t \RR$, namely
 \begin{multline} \label{W}
 W_t \RR = K_t \RR + \\ \sum_{n=1}^\infty \, (-1)^n \int_{t > t_1 >t_2 > \cdots > t_n} \, \left [ K_{t-t_1} B K_{t_1 -t_2} B \cdots K_{t_n} \right ] \RR
 \end{multline} 
 where 
 \begin{equation}
 K_t \RR = \left [ e^{-t \delta^2 H_0} \right ] \RR .
 \end{equation}
 
 In the standard case $H_0= - \Delta$, one can check that the series (\ref{W}) corresponds to an asymptotic expansion in $\delta$, when we use 
 cartesian coordinates.
 The interest of this approach lies however in the fact that we can treat systems where more general coordinate systems are useful and particularly when
 the potential $V$ contains a Coulomb potential, for which none of the standard techniques could work, to the best of our knowledge.
 
 \section{Potential with a cylindrical symmetry and parabolic coordinates}
 
 We will consider the case of a three dimensional system with a potential $V$ invariant under the rotation around the $z$-axis. We can therefore stay
 in the subspace corresponding to $L_z=m$. In what follows, $m$ will be fixed, even when $\hbar \To 0$. It is appropriate to use the following parabolic
 coordinates:
 \begin{equation} \label{coordinates}
 \left \{
          \begin{array}{l} X = \sqrt{xy} \cos{\phi} \\
                           Y = \sqrt{xy} \sin{\phi} \\
                           Z = \frac{1}{2} \left ( x-y \right)
          \end{array} \right.
 \quad  \quad x,y \geq 0 
 \end{equation}
 in which a Coulomb potential $\mu r^{-1}$ becomes $\frac{2 \mu}{x+y}$.
 
 The dimensionless hamiltonian we will consider are of the form, 
 \begin{equation}
 H = - \delta^2 \Delta_m + \mathcal{V} - \frac{\mu}{r}
 \end{equation}
 where the potential $V$ has been decomposed into a Coulomb part and a more regular one $\mathcal{V}$.
 
 The eigenvalue equation $H\psi = E \psi$ will be written as 
 \begin{equation} \label{eigenvalue relation}
 \left [ \delta^2 A + \VT \right ] \psi = \mu \psi
 \end{equation}
 where 
 \begin{equation} \label{A}
 A = h_x + h_y ,
 \end{equation}
 \begin{equation}
 h_x = - \frac{d}{dx} x \frac{d}{dx} + \frac{m^2}{4x} + \frac{\epsilon}{2 \delta^2} x
 \end{equation}
 and
 \begin{equation} \label{VT}
 \VT(x,y) = \left [ \mathcal{V} - E_0 \right ] \frac{x + y}{2}.
 \end{equation}
 
 Here $\epsilon = E_0 - E > 0 $, $E_0$ being the possible ionization threshold, which may depend on $\delta$.
 We now consider that equation (\ref{eigenvalue relation}) quantizes $\mu$ into eigenvalues $\mu_j(E)$ when $\epsilon>0$. 
 If we define the partition function 
 \begin{equation}
 \mathcal{G} = \text{Tr }e^{-t \left [ \delta^2 A + \VT \right ]} = \sum_j \, e^{-t \mu_j}
 \end{equation}
 and note that $\frac{d \mu_j}{dE} = - \left (\psi_j , \, \frac{x+y}{2} \psi_j \right ) < 0$ which implies 
 \begin{equation}
 \Theta \left ( \mu - \mu_j \right ) = \Theta \left ( E -E_j(\mu) \right ),
 \end{equation}
 and finally the partition function
 \begin{equation}
 \mathcal{G} = \int d\mu \, e^{-t \mu} \, \frac{d}{d\mu} \sum_j \, \Theta \left ( E -E_j(\mu) \right ).
 \end{equation}
 Hence the density of states we are looking at 
 \begin{equation}
 \mathcal{N}(E,\mu) = \sum_j \, \Theta \left ( E -E_j(\mu) \right )
 \end{equation}
 where $\mu$ is given, can be obtained once we know the partition function $\mathcal{G}$. \\
 But
 \begin{equation}
 \mathcal{G} = \iint dx dy \, U_t \XY
 \end{equation}
 so that our task consists now in computing perturbatively $\mathcal{G}$ using equations (\ref{eigenvalue relation}), (\ref{A}) and (\ref{VT}). 
 For this purpose we need to compute 
 \begin{equation}
 K_t \xy = \left [e^{-t \delta^2 A } \right ] \xy
 \end{equation}
 This can be done by using the fact that the eigenfunctions of $\delta^2 h_x$ are $\varphi_n\left(\frac{x}{\gamma}\right)$ where 
 \begin{equation}
 \varphi_n (x) = x^{\left \vert \frac{m}{2} \right \vert} \, e^{-\frac{x}{2}} \, L_n^{\vert m \vert} (x) \sqrt{\frac{n!}{(n+m)!}}
 \end{equation}
 $L_n^{\vert m \vert} (x)$ being a Laguerre polynomial and $\gamma = \frac{\delta}{\sqrt{2\epsilon}}$. The final result can be expressed as
 \begin{widetext}
 \begin{multline} \label{KT}
 K_t \xy = \frac{\theta \, e^{-4 \vert m \vert \theta} }{4 \pi t \delta^2 \sinh{\theta} \sqrt[4]{x x' y y'}}\,
 \exp{\left ( - \frac{\theta}{t \delta^2 \tanh{(\theta)}} \left [ (\sqrt{x} - \sqrt{x'})^2 + (\sqrt{y} - \sqrt{y'})^2 \right ] \right )} \\
 \exp{\left ( - \frac{2 \theta \tanh{\left (\frac{\theta}{2} \right )}}{t \delta^2} \left [ \sqrt{x x'} + \sqrt{y y'} \right ] \right ) } \,
 \hat{I}_{m} \left ( \frac{2 \theta}{t \delta^2 \sinh{(\theta)}} \sqrt{x x'}\right ) \hat{I}_{m} \left ( \frac{2 \theta}{t \delta^2 \sinh{(\theta)}}
  \sqrt{y y'}\right )
 \end{multline}
 where $\theta = \frac{t \delta \sqrt{2 \epsilon}}{2}$ and we introduced for later purpose the function
 \begin{equation}
 \hat{I}_m (x) = \sqrt{2 \pi x} \, e^{-x} \, I_m(x)
 \end{equation}
 $I_m(x)$ being the usual Bessel function. Its usefulness comes from its simple asymptotic expansion
 \begin{equation}
 \hat{I}_m (x) = 1 - \frac{1}{2x} \left (m^2 - \frac{1}{4} \right ) + \mathcal{O} \left ( \frac{1}{x^2} \right ).
 \end{equation}
 
 The semi-classical limit of the propagator $K_t$ that we will denote by $\hat{K}_t$ is obtained by taking $\delta \To 0$, $ \vert m \vert \delta \To 0$ 
 and $(x,x',y,y')$ fixed
 \begin{equation}
  \hat{K}_t \xy = \frac{\exp{\biggl ( - \frac{1}{t\delta^2} \left [ (\sqrt{x} - \sqrt{x'})^2 + (\sqrt{y} - \sqrt{y'})^2 \right ] - 
  \frac{\epsilon t}{4} \left ( \sqrt{x x'} + \sqrt{y y'} \right ) \biggr )}}{4 \pi t \delta^2 \sqrt[4]{x x' y y'}}
 \end{equation}
\end{widetext}
\section{The partition function}
We will compute $U_t$ and therefore $\mathcal{G}$ up to order $\delta^0$. We decompose 
\begin{equation}
U_t = U^0_t + U^1_t + U^2_t
\end{equation}
using the simplifying notation $r= (x,y)$,
\begin{equation}
U^0_t \RR = e^{-t \TV (r)} \, K_t \left ( r \vert r \right ) ,
\end{equation}
\begin{widetext}
\begin{equation}
U^1_t \RR = e^{-t \TV (r) } \, \int_{t >t_1>0} \, K_{t-t_1} \left ( r \vert r' \right ) K_{t_1} \left ( r' \vert r \right ) \left [ \TV(r')- \TV(r) \right ] dr',
\end{equation}
\begin{equation} \label{U2}
U^2_t \RR = e^{-t \TV (r) } \, \int_{t >t_1>t_2>0} \, K_{t-t_1} \left ( r \vert r_1 \right ) K_{t_1-t_2} \left ( r_2 \vert r \right ) \left [ \TV(r_1) - \TV(r) \right ] \left [ \TV(r_2) -\TV(r) \right ].
\end{equation}
Let us begin by computing $U^1_t$ to order $\delta^0$. For this purpose we replace $K_t$ by $\hat{K}_t$.
\begin{equation}
K_{t-t_1} \RR K_{t_1} \RR =  \frac{\exp{ \biggl ( -\frac{\epsilon}{4} t \left [ \sqrt{x x'} + \sqrt{y y'} \right ] \biggr ) }}{(4 \pi)^2 \delta^4 \sqrt{x y x' y'} (t-t_1) t_1} \exp{\Biggl ( -\frac{\left ( \sqrt{x} - \sqrt{x'} \right )^2 + \left ( \sqrt{y} - \sqrt{y'} \right )^2}{2 \bar{\delta}^2} \Biggr )}
\end{equation}
where 
\begin{equation}
 \frac{\delta^2}{\bdelta^2} = \frac{2t}{(_{}t-t_1) t_1}.
\end{equation}

Let us now make the change of variables
\begin{align}
 \sqrt{x'} &= \sqrt{x} + \bdelta u \\
 \sqrt{y'} &= \sqrt{y} + \bdelta v
\end{align}
with $\bdelta$ being small then
\begin{equation} \nonumber
\TV(r') - \TV(r) = 2 \bar{\delta} \left [ \TV_x \sqrt{x} u + \TV_y \sqrt{y} v \right ] + 2 \bar{\delta}^2 \left [ \TV_x \frac{u^2}{2} + \TV_{xx} x u^2 + \TV_y \frac{v^2}{2} + \TV_{yy} y v^2 + 2 \TV_{xy} \sqrt{x y} u v \right ] + \mathcal{O}\left ( \bar{\delta}^3 \right )
\end{equation}
where we have denoted by $\TV_x \equiv \partial_x \TV$.
Introducing in addition
\begin{equation}
 \TTV = \TV + \epsilon \frac{x+y}{2},
\end{equation}
it yields 
\begin{align}
U^1_t (x, y) &= \frac{e^{-t \TTV }}{2 \pi^2 \delta^4 \sqrt{x y}} \int_0^t \, \frac{dt_1}{(t-t_1) t_1}  \bdelta^3 \iint \, du dv \, \exp{ \left ( -\frac{u^2 + v^2}{2} - \frac{\epsilon \bdelta}{2} t \left ( \sqrt{x} u + \sqrt{y} v \right ) \right ) } \nonumber \\
& \quad \times  \Biggl \{ \left [ \TV_x \sqrt{x} u + \TV_y \sqrt{y} v \right ] + \bdelta \left [ \TV_x \frac{u^2}{2} + \TV_{xx} x u^2 + \TV_y \frac{v^2}{2} + \TV_{yy} y v^2 + 2 \TV_{xy} \sqrt{x y} u v \right ] \Biggr \} \nonumber \\
&= \frac{2 \pi \delta^4 e^{-t \TTV }}{2 \pi^2 \delta^4 \sqrt{x y}} \int_0^t dt_1  \, \frac{(t-t_1)^2 t_1^2}{4t^2(t-t_1) t_1} \left [ - \frac{\epsilon t}{2} \left ( \TV_x x + \TV_y y \right ) + \frac{1}{2} \left ( \TV_x + \TV_y \right ) + \TV_{xx} x + \TV_{yy} y \right ] 
\end{align}
so that finally one finds for $U_t^1$
\begin{equation}
    U^1_t \left ( x, y \right ) = - \frac{e^{-t \TTV}}{48 \pi \sqrt{x y}} \, t \, \left [ \TV_x \left ( 1- \epsilon t x \right ) + \TV_y \left ( 1- \epsilon t y \right ) + 2 \TV_{xx} x + 2 \TV_{yy} y \right ] + \mathcal{O}(\delta)
\end{equation}
\end{widetext}
It is worth noting that we have replaced the domain of integration in the variable $u$: $u \geq -\frac{\sqrt{x}}{\delta}$ by $u \geq - \infty$. This accounts to neglect terms of the order $\exp{\left ( - \frac{t}{\delta} \right)}$.
Such terms ( essential singularities in $\delta$ ) are responsible for the oscillating terms in the density of states, related to the classical periodic orbits.

The second term $U_t^2 (x, y)$ is treated in a similar way, replacing again $K_t$ by $\hat{K}_t$ in equation (\ref{U2}), one finds
\begin{equation}
 U_t^2 (x, y) = \frac{t^2 \,e^{-t\TTV }}{48 \pi \sqrt{x y}} \left [ \TV_x^2 x + \TV_y^2 y \right ].
\end{equation}

These two terms give a contribution to the partition function, that we denote by $\mathcal{G}^1$
\begin{equation}
 \mathcal{G}^1 = \iint dx dy \, \left [ U_t^1 (x, y) + U_t^2 ( x, y) \right ].
\end{equation}
It remains to compute the part of the partition function that we will denote by $\mathcal{G}^0$
\begin{equation}
 \mathcal{G}^0 = \iint dx dy \, e^{-t \TV (x, y)} K_t \left ( x y \vert x y \right).
\end{equation}
\begin{widetext}
Using equation(\ref{KT}), it is given more explicitly by 
\begin{equation}
 \mathcal{G}^0 = \mathcal{C} \iint \frac{dx dy}{\sqrt{xy}} \exp{\left ( -t \TV(x, y) - \frac{x + y}{\gamma \sinh{(\theta)}} \left ( \cosh{(\theta)} - 1 \right ) \right ) } \IM \left ( \frac{x}{\gamma\sinh{(\theta)}} \right ) \IM \left ( \frac{y}{\gamma\sinh{(\theta)}} \right )
\end{equation}
where 
\begin{equation*}
 \mathcal{C} = \frac{e^{- 4 \theta \vert m \vert}}{8 \pi \gamma \sinh{(\theta)}} \quad \textrm{ and } \quad \gamma = \frac{\delta}{\sqrt{2 \epsilon}}.
\end{equation*}
Using the identity
\begin{equation}
 \IM (a) \IM (b) = 1 + \left ( \IM(a) -1 + \IM (b) -1 \right ) + \left ( \IM(a) -1 \right ) \left ( \IM(b) -1 \right ),
\end{equation}
we decompose $\mathcal{G}^0$ into three different parts that we treat differently. 
A typical term will be of the form
\begin{equation}
 A = \int_0^\infty \frac{dx}{\sqrt{x}} \left [ \IM\left ( \frac{x}{\gamma \sinh{(\theta)}} \right ) -1 \right ] \exp{\left ( - \frac{x}{\gamma \sinh{(\theta)}}\left ( \cosh{(\theta)} -1 \right ) \right )} g(x).
\end{equation}
We decompose $A$ as follow: $A = A_1 + A_2$ 
\begin{equation}
 A_1 = g(0) \int_0^\infty \frac{dx}{\sqrt{x}} \left [ \IM\left ( \frac{x}{\gamma \sinh{(\theta)}} \right ) -1 \right ] \exp{\left ( - \frac{x}{\gamma \sinh{(\theta)}}\left ( \cosh{(\theta)} -1 \right ) \right )},
\end{equation}
\begin{equation}
 A_2 = \int_0^\infty \frac{dx}{\sqrt{x}} \left [ \IM\left ( \frac{x}{\gamma \sinh{(\theta)}} \right ) -1 \right ] \exp{\left ( - \frac{x}{\gamma \sinh{(\theta)}}\left ( \cosh{(\theta)} -1 \right ) \right )} \bigl ( g(x) - g(0) \bigr ).
\end{equation}
\end{widetext}
But 
\begin{equation}
 A_1 = g(0) \sqrt{2 \pi \gamma \sinh{(\theta)}} \left ( \frac{e^{-\theta \vert m \vert}}{\sinh{(\theta)}} - \frac{1}{2 \sinh{\left ( \frac{\theta}{2}\right )}} \right )
\end{equation}
and
\begin{equation}
 \lim_{\delta \rightarrow 0} \frac{A_2}{\gamma \sinh{(\theta)}} = - \left ( \frac{m^2}{2} - \frac{1}{4} \right ) f(x)
\end{equation}
where 

\begin{equation}
 f(x) = \int_0^\infty \frac{dx}{x^{3/2}} e^{-\frac{\epsilon t x}{2}} \bigl [ g(x) - g(0) \bigr ] .
\end{equation}

In order to obtain this result, we have replaced $\IM(x)$ by its asymptotic behavior. In this way we obtain if $\mathcal{G}^0 = \mathcal{G}^0_0 + \mathcal{G}^0_1 + \mathcal{G}^0_2$
\begin{widetext}
\begin{equation}
 \mathcal{G}_0^0 = \left [ \frac{1}{4 \pi t \delta^2} - \frac{\vert m \vert}{2 \pi \delta}\sqrt{2 \epsilon} + \frac{\epsilon t}{\pi} \left ( m^2 - \frac{1}{48} \right )  \right ] \iint \frac{dx dy}{\sqrt{xy}} e^{- t \TV(xy)}  + \frac{\left ( \epsilon t \right )^2}{4 \cdot 48 \pi } \iint \frac{dx dy}{\sqrt{x y}} e^{-t \TTV (xy)} \left ( x + y \right ),
\end{equation}
\begin{multline}
 \mathcal{G}^0_1 = \left [ -\frac{\vert m \vert}{4 \delta \sqrt{\pi t}} + \sqrt{\frac{\epsilon t}{2 \pi}} \left (m^2 - \frac{7}{32} \right ) \right ] g(0) + \frac{ m^2 - \frac{1}{4}}{8 \pi } \epsilon t \iint \frac{dx dy}{\sqrt{x y}} e^{-t \TTV (xy)} \\ + \frac{t}{8 \pi} \left ( m^2 - \frac{1}{4} \right ) \iint \frac{dx dy}{\sqrt{x y}} e^{-t \TTV(xy)} \left [ \TV_x + \TV_y \right ],
\end{multline}
\begin{equation}
 \mathcal{G}^0_2 = \frac{m^2}{4} e^{-t \TTV(0,0)}
\end{equation}
where $g(x)= \int \frac{dy}{\sqrt{y}} \, e^{-t \VT(x,y) - \frac{\epsilon t}{2} y}$.
\end{widetext}
In deriving these expressions we have implicitly used some assumptions about the potential $V(x,y)$.
All the expressions given are correct provided 
\begin{align}
 V(x,y) - V(0,y) &\sim x^\alpha \quad \textrm{ when } x \To 0 \\
 V(x,y) - V(x,0) &\sim y^\alpha \quad \textrm{ when } y \To 0
\end{align}
and $\alpha + \frac{1}{2} > 0$.

\section{The integrated density of states}

The integrated density of states (IDOS) $\mathcal{N}(E) = \sum_j \Theta \left (E -E_j \right )$ can now be obtained from the knowledge of the partition function. But we must not forget that the ionization threshold $E_0$ may also depend on $\delta$. We assume that $E_0 = \delta e_1$ and consider that semi-classically the IDOS should be computed by keeping $\epsilon = E -E_0$ fixed and taking $\delta$ small.

We therefore take
\begin{equation}
 \TTV(x,y) = W(x,y) - \delta e_1 \frac{x+y}{2}
\end{equation}
where $W(x,y) = \frac{x+y}{2} \left ( \mathcal{V}(x,y) + \epsilon \right )$.
We decompose now the IDOS into three parts corresponding to their importance in the semi-classical limit
\begin{equation} \label{N}
 \mathcal{N} = \frac{N_0}{\delta^2} + \frac{N_1}{\delta} + N_2.
\end{equation}
Using the representation 
\begin{equation}
 \frac{1}{t} e^{-t a} = \int d\mu \, e^{-t \mu} \frac{d}{d\mu} \left [ \mu - a \right ]_{+}
\end{equation}
we see that
\begin{equation}
 \mathcal{N}_0 = \frac{1}{4\pi} \iint \frac{dx dy}{\sqrt{xy}} \left [ \mu - W(x,y) \right ]_{+}.
\end{equation}
If we go back to the original coordinates $(\rho, z)$ instead of the parabolic ones, we can rewrite this term in a more physical expression
\begin{equation}
 \mathcal{N}_0 = \left ( \frac{1}{2\pi} \right )^2 \int d^2p \int d\rho dz \, \Theta \left ( - \epsilon - \frac{p^2}{2} - \mathcal{V}(\rho, r) + \frac{\mu}{r} \right )
\end{equation}
which is the Thomas-Fermi form of the density of states, except that the volume element is $d\rho dz$, not $\rho d\rho dz$ and $- \epsilon$ replaces $E$.
In  order to compute the other terms, we can use the identities
\begin{align}
 e^{-t a} &= \int d\mu \, e^{-t \mu} \frac{d}{d\mu} \Theta \left ( \mu - a \right ) \\
\frac{1}{\sqrt{t}} e^{-t a} &= \int d\mu \, e^{-t \mu} \frac{d}{d\mu} \frac{2}{\sqrt{\pi}} \left [ \mu - a \right ]_{+}^{\frac{1}{2}}
\end{align}
We then found that
\begin{multline}
 \mathcal{N}_1 = -\frac{\vert m \vert}{2 \pi} \sqrt{2\epsilon} \iint \frac{dx dy}{\sqrt{x y}} \Theta \left ( \mu - W(x,y) \right )  \\ + \frac{e_1}{8 \pi} \iint \frac{dx dy}{\sqrt{x y}} \Theta \left ( \mu - W(x,y) \right ) \left ( x + y \right ) - \frac{\vert m \vert \mu}{\sqrt{2 \epsilon}}
\end{multline}
since $W(0,y)=W(x,0)=0$ by our assumptions and
\begin{equation}
 \mathcal{N}_2 = \frac{1}{4} \left ( 9 m^2 - \frac{7}{4} \right ) + \frac{1}{\pi} \iint \frac{dx dy}{\sqrt{xy}} \delta \left ( \mu - W(x,y) \right ) g(x,y)
\end{equation}
where 
\begin{align}
 g(x,y) &= \epsilon m^2 - \vert m \vert \sqrt{\frac{\epsilon}{2}} e_1 \frac{x+y}{2} \nonumber \\
 & \quad + \frac{e_1^2}{8} \left ( \frac{x +y}{2} \right )^2 + \frac{1}{8} \left ( m^2 - \frac{1}{3} \right ) \left ( W_x + W_y \right ) \nonumber \\  & \quad-\frac{1}{48} \left ( x W_{xx} + y W_{yy} \right )
\end{align}

\begin{figure*}
  \includegraphics{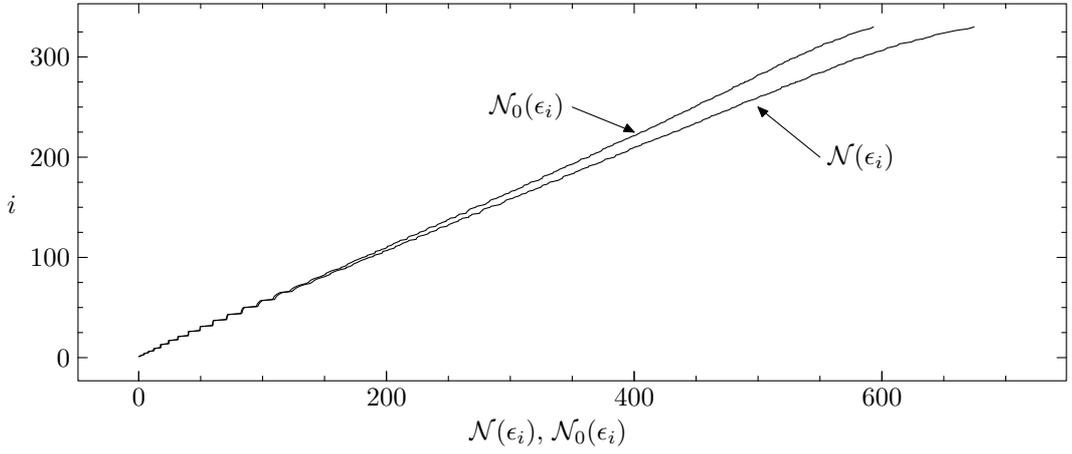}
  \caption{Plots of numerical energy counting versus the semi-classical integrated density of states for $\delta = 4.64 \cdot 10^{-2}$ ($\gamma=10^{-4}$), once the Thomas-Fermi approximation $\mathcal{N}_0(\epsilon_i)$ and then with corrections up to second order $\mathcal{N}(\epsilon_i)$.}
  \label{state density 1}
\end{figure*}
\begin{figure*}
  \includegraphics{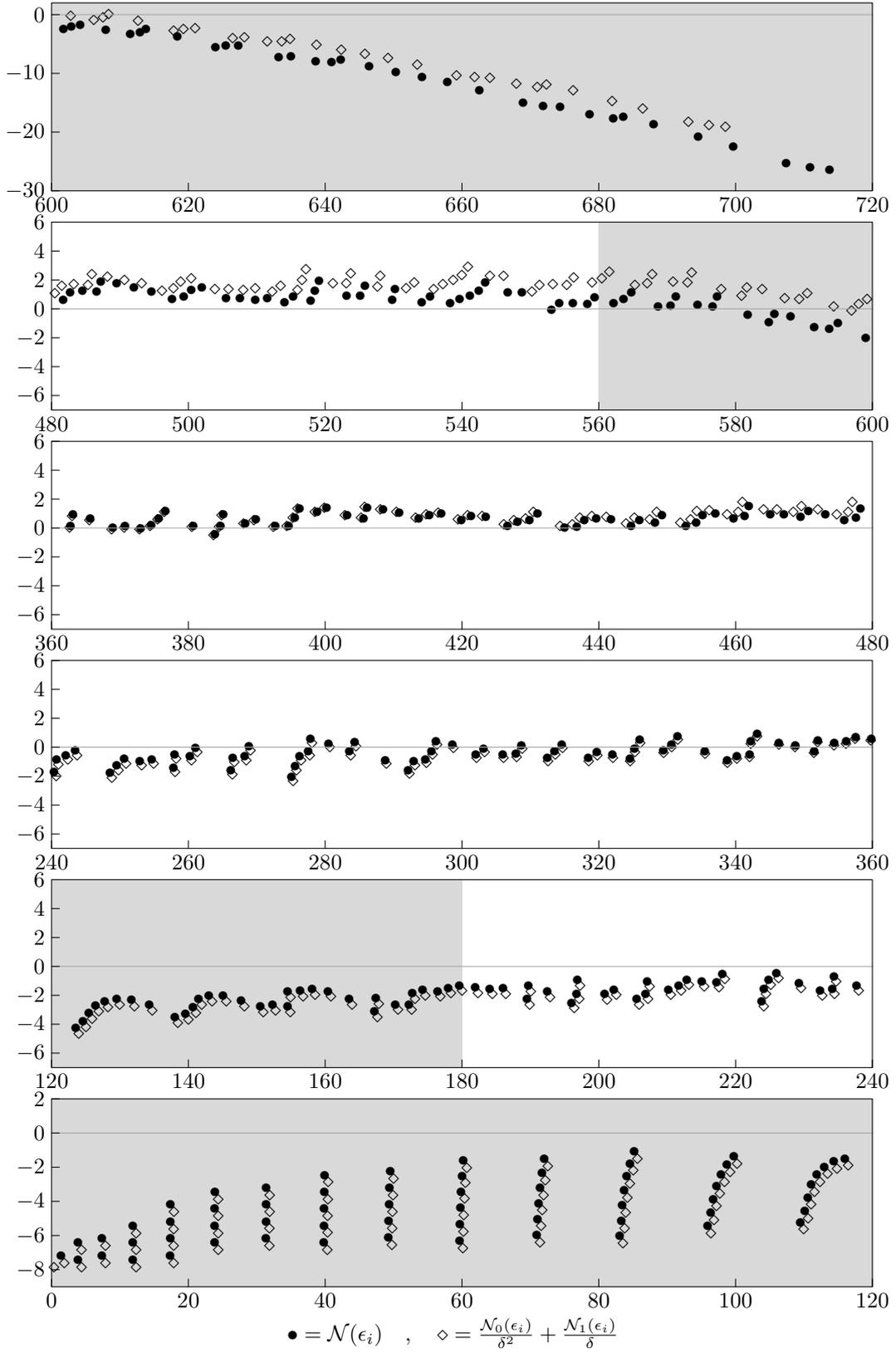}
  \caption{Difference between the semi-classical integrated density of states for $\delta = 4.64 \cdot 10^{-2}$ ($\gamma=10^{-4}$) and its best fit curve $y=\frac{1}{2}x+h$ computed in the non-shadowed region $i \in [90;300]$. $h$ is equal to $8.56$ in the case of first order corrections only and $8.39$ in the case of both first and second order corrections. The existence of a quasi-invariant at low energies is clearly distinguished by the vertical clustering produced by similar energy values.}
  \label{state density 2}
\end{figure*}

\section{The integrated density of states of the hydrogen atom in a uniform magnetic field} \label{The density of states}

The case of the hydrogen atom in a magnetic field constant and directed along the $z$ axis represents the most important application of our general formula.

In term of the units of length $a_0 = \frac{\hbar^2}{M e^2}$ and $L = \sqrt[3]{\frac{M}{B^2}}$, our parameter $\delta$ will be $\delta=\sqrt{\frac{a_0}{L}}$. Here $M$ is the reduced mass, $B$ the magnetic field strength and $e$ the charge of the electron. In units of energy $\frac{e^2}{L}$, the dimensionless hamiltonian reads
\begin{equation}
 H = -\frac{\delta^2}{2} \Delta_m + \frac{1}{8} \rho^2 + \frac{\delta m}{2} - \frac{1}{\sqrt{\rho^2 + z^2}},
\end{equation}
the parameter $\epsilon$ is given by 
\begin{equation}
\epsilon = \frac{\delta m}{2} + \delta e_1 - E  
\end{equation}
where $ e_1 = \frac{1}{2} ( 1 + \vert m \vert )$ and finally the diamagnetic potential in this case is expressed as
\begin{equation}
 \mathcal{V} (x,y) =\frac{x y}{8}.
\end{equation}
We may note that many authors \cite{Clark1980, Clark1982, Gay1983, Delande1984, Delande1986, Delande1991, Delande1991a, Friedrich1989, Wintgen1986, Wintgen1986a, Wintgen1986b, Wintgen1986c, Wintgen1986d, Wunner1985, Wunner1985a, Wunner1986} have introduced instead of our parameter $\delta$ a parameter $\gamma=\delta^3= \frac{eB}{\hbar}a_0^2$.\\
Let us use the parametrization $c \in [1, \infty]$ instead of $\epsilon \in [0, \infty]$ given by the relation
\begin{equation}
 2 \epsilon = (c-1) c^{-\frac{2}{3}},
\end{equation}
in this case the various terms in the IDOS become 
\begin{widetext}
\begin{align} \label{fresults1}
 \mathcal{N}_0(\epsilon) &= \frac{1}{c^{4/3} \pi} \int_0^1 \frac{dy}{\sqrt{y}} \left ( 1 + \frac{ 2y}{c-1 + y} \right ) \sqrt{R(y)}, \nonumber \\
\mathcal{N}_1(\epsilon) &= - \vert m \vert \frac{\sqrt[3]{c}}{\sqrt{c-1}} - \frac{2 \vert m \vert}{\pi} \sqrt[3]{c} \int_0^1 \frac{dy}{\sqrt{y}} \left ( 1 + \frac{ 2y}{c-1 + y} \right ) \frac{1}{\sqrt{R(y)}} + \frac{1}{\pi} (1 + \vert m \vert ) c^{-\frac{2}{3}} \int_0^1 \frac{dy}{\sqrt{y}} \frac{\sqrt{R(y)}}{c-1 + y} \nonumber, \\
\mathcal{N}_2(\epsilon) &= \frac{1}{4} \left ( 9 m^2 - \frac{7}{4} \right ) + \int_0^1 \frac{dy}{\sqrt{y}} \frac{f(y)}{\sqrt{R(y)}},
\end{align}
\end{widetext} where we have defined for ease the functions
\begin{equation}\label{fresults2}
 R(y) = \left ( 1-y \right ) \left ( y^2 + (2c-1)y + c^2 \right )
\end{equation}
and
\begin{equation} \label{fresults3}
 f(y) = a_1 + \frac{a_2}{c-1+y} + a_3 y + \frac{a_4}{(c-1+y)^2}
\end{equation}
 together with the coefficients
\begin{align} \label{fresults4}
 a_1 &= \frac{c-1}{4\pi} \left ( 9 m^2 - \frac{1}{3} \right ), \\
 a_2 &= - \frac{2}{\pi} \vert m \vert \left ( 1 + \vert m \vert \right ) c \sqrt{c-1} , \\
 a_3 &= \frac{1}{2\pi} \left ( \frac{m^2}{2} - \frac{1}{3} \right ) ,\\
 a_4 &= \frac{c^2}{\pi} \left ( m^2 + \vert m \vert + \frac{1}{3} \right ) .
\end{align}
All these integrals are of the elliptic type and thus evaluations have been performed numerically. \\
It is important to notice that the Thomas-Fermi IDOSs remains finite at the ionization threshold $(\epsilon=0)$. This is however not compatible with the diverging IDOS at this threshold as it corresponds to an accumulation point for the eigenenergies. We therefore expect that the corrections $\mathcal{N}_1(\epsilon)$ and  $\mathcal{N}_2(\epsilon)$ become more and more  relevant in the limit $\epsilon \rightarrow 0$. This will be confirmed in the next section by the numerical evaluation of the IDOS.

\section{Comparison with numerical results}

\begin{figure*}
  \includegraphics[width=\textwidth]{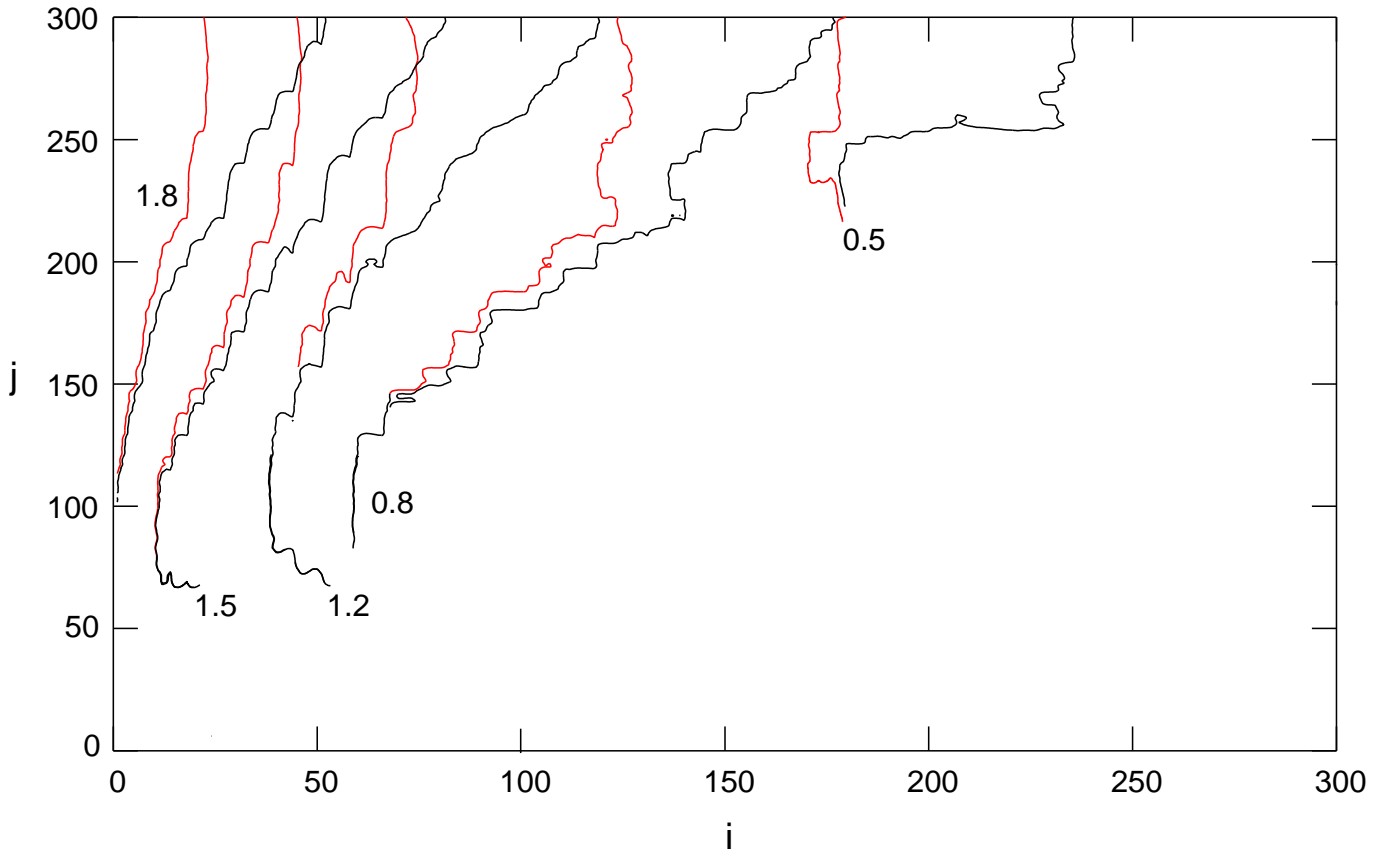}
  \caption{Contourplot for $\delta = 4.64 \cdot 10^{-2}$ ($\gamma=10^{-4}$) of the RMS of residuals between the semi-classical integrated density of states and the theoretical curve given by $y=\frac{1}{2} x + h$ where the height $h$ is left as a free parameter within the $\epsilon_k$ interval given by $k \in [i,j]$. The black curve corresponds to a calculation with first order correction only, the gray one with both first and second order corrections. }
  \label{state density 3}
\end{figure*}

The preceding formulas have been compared with numerical results of the energy levels in the $m^\pi = 0^+$ subspace of the hydrogen atom in a magnetic field strength of $23.5\, T$ that is $\delta=4.64 \cdot 10^{-2}$ ($\gamma=10^{-4}$).
The latter were obtained through a high precision finite element approach in cylindrical coordinates. Details of the numerical method and its comparison with other approaches and results will be given in forthcoming paper \cite{schuepbach2007}.

Choice has been made to present the results as a parametric function $(\mathcal{N}(\epsilon_i); i)$ where $\epsilon_i$ correspond to the $i^{th}$ numerical energies. Nevertheless with regards to the subspace considered, only even eigenfunctions are taken into account. Therefore we expect a linear correspondence with a $1/2$ slope since, on average between each even energy lies an odd one.\\
Figure \ref{state density 1} illustrates the parametric functions obtained through the Thomas-Fermi IDOS and its first and second order corrections with regards to $\delta$. Both functions exhibit a clear linear growth as expected. However there are increasing deviations of the Thomas-Fermi curve at high energies corroborating our last remarks in section \ref{The density of states}, namely that the Thomas-Fermi term is not valid as one approaches the ionization threshold.

A magnification of the linear correspondance to the theoretical half slope is presented in Figure \ref{state density 2} where deviations from the best fit line $y=\frac{1}{2}x + h$ are reported once for the function $\mathcal{N}(\epsilon)$ and once without the second order correction i.e. $\mathcal{N}(\epsilon) - \mathcal{N}_2(\epsilon)$. $h$ is left as a free parameter since its value cannot be inferred rigorously in the analytical approach (no absolute counting from below). As expected, one notices failure to match a linear approximation in the shaded regions corresponding to the low and very high energy range. Strong deviations are indeed observed in the former domain due to the existence of a quasi-invariant which generates quasi-degeneracies. \\
The magnetic field is here not strong enough to take over the Coulomb term and one observes clearly on Figure \ref{state density 2} a clustering of the eigenvalues around the quasi-degeneracies for which, at the lower end, simple perturbation theory in the magnetic field strength would already provide an accurate picture. At the higher end of the lowest shaded region one would probably benefit from an analytical expression of the hamiltonian in the Gay-Delande's basis \cite{Gay1983}, and then apply a semi-classical expansion in order to account for the density of
states. \\
The deviations at high energies can be attributed to two presumed causes, namely missing higher correction orders or the departure from the semi-classical domain. However further analysis would be required in order to determine which of them is the first dominant.

Although in the mid domain deviations from the theoretical fit line are small, it is not immediately clear how one could define the best domain boundaries for each curve. Moreover those boundaries are required in order to compute the $h$ parameter.
In order to resolve such ambiquities we plot in Figure \ref{state density 3} the RMS of residuals around linearity $y = \frac{1}{2} x + h$ with $h$ left as a free parameter for every possible intervals $\epsilon_k$, $k \in [i,j]$. Contributions arising from the first and second order corrections have been computed separately so that one can easily pictures the difference. 
The results show indeed that including first and second order corrections improves the overall linearity in different ways. One can think first at fixed interval where the RMS of residual would then be lower or second at constant RMS residual where the range would then be wider. The latter is particularly important as it clearly demonstrates the tiny refinement induced by the second order corrections at high energies. \\
Considering the non-shadowed domain of Figure \ref{state density 2}, the RMS of residuals equals $1.06$ with all corrective terms and $1.35$ with just the first order correction.

It is now quite interesting to compare the different contributions of $\mathcal{N}(\epsilon)$ represented by $\mathcal{N}_0$, $\mathcal{N}_1$ and $\mathcal{N}_2$ in Eq.~(\ref{N}) to the slope of figure \ref{state density 1}. For this purpose, the energy interval is taken from the preceding considerations, namely $\epsilon_k$, $k \in [90,280]$. The exact numerical values of the best fit line are presented in table \ref{table 1}. It appears clearly again that the Thomas-Fermi term i.e. $\delta^{-2}\mathcal{N}_0$, is definitively not able to reproduce quantitavely the counting function. A very important contribution is indeed brought by the first order correction $\mathcal{N}_1$ in terms of accuracy with respect to the theoretical slope value. At last, the second order correction $\mathcal{N}_2$ further improves the slope value but with higher RMS of residuals. 
The lower RMS of residuals value when ignoring $\mathcal{N}_2$ is an artefact since it was measured with respect to a less correct slope. Figure \ref{state density 3} gives a more consistant picture on the issue of analyzing the RMS fluctuations.
In fact, we would like to stress that fluctuations of the energy level spacings are an important feature of the spectrum of the hydrogen atom under constant magnetic field, and carry out important statistical informations related to quantum chaos \cite{Berry1977, Bohigas1984, Delande1986, Wunner1986, Zakrzewski1995, Stoeckmann1999, Haake2001}.They will be analyzed in a forthcoming paper \cite{schuepbach2007} since these are out of the scope of the present paper.

\begin{table}[t]
\vspace{.3cm}
 \begin{tabular}{c|ccc}
function & slope & height & rms of residuals\\
\hline \\
$\frac{\mathcal{N}_0(\epsilon)}{\delta^2} $ & $0.5649 \pm 0.0007$ & $-3.53 \pm 0.25 $ & $0.98$\\ 
$\frac{\mathcal{N}_0(\epsilon)}{\delta^2}+ \frac{N(\epsilon)}{\delta}$ & $0.5117 \pm 0.0003$ & $4.43 \pm 0.12 $ & $0.48$\\ 
$\mathcal{N}(\epsilon)$ & $0.5084 \pm 0.0004$ & $5.43 \pm 0.13$ & $0.54$ \\ 

\end{tabular}
\caption{Best linear least square approximations of the smoothed density of states in the energy range starting at the $90^{th}$ even energy levels up to the $280^{th}$.} 
\label{table 1}
\end{table}

\noindent Finally a last remark: it should be clear to the reader that both first and second order corrections do depend on the magnetic quantum number $m$. Although we have only computed the case $m=0$, corrections to the Thomas-Fermi term would be higher for $m\neq 0$ according to Eqs. (\ref{fresults1}-\ref{fresults4}). 

\section*{Acknowledgments}
This work was supported by the "Fond National Suisse de la Recherche Scientifique".

\bibliography{dks}

\end{document}